\documentclass[a4paper,11pt]{article}

\usepackage{dynkin-diagrams}
\pgfkeys{/Dynkin diagram, edge length=1cm,
fold radius=.6cm, 
root-radius=.11cm,
indefinite edge/.style={
draw=black, fill=white, dotted,
thin}}

\usepackage{xypic}
\usepackage{tikz-cd}

\usepackage{pos}

\def\nn{\nonumber}

\def\fraction#1{\small{1\over#1}}
\def\fr{\fraction}

\def\fg{{\mathfrak g}}

\def\fk{\mathfrak{k}}

\def\so{{\mathfrak{so}}}
\def\sl{{\mathfrak{sl}}}
\def\gl{{\mathfrak{gl}}}

\def\RR{{\mathbb R}}

\def\ZZ{{\mathbb Z}}

\def\LL{{\mathscr L}}

\def\*{\partial}

\def\lbr{[\hspace{-2.25pt}[}
\def\rbr{]\hspace{-2.25pt}]}

\def\eg{\hbox{\it e.g.}}
\def\ie{\hbox{\it i.e.}}

\title{BV actions for extended geometry}

\author*[a]{Martin Cederwall}

\affiliation[a]{Dept. of Physics, Chalmers Univ. of Technology,\\ 
  SE-412 96 Gothenburg, Sweden}

\emailAdd{martin.cederwall@chalmers.se}

\abstract{I review the construction of actions for extended geometry from the grading of an underlying tensor hierarchy algebra, which provides the full set of Batalin--Vilkovisky fields. The dynamics is neatly encoded in a complex. This talk, presented at the Corfu Summer Institute 2024, is mainly based on joint work with J. Palmkvist, in particular ref. \cite{Cederwall:2023xbj}.}

\FullConference{Proceedings of the Corfu Summer Institute 2024 "School and Workshops on Elementary Particle Physics and Gravity" (CORFU2024)\
12 - 26 May, and 25 August - 27 September, 2024\\
Corfu, Greece\\}

\tableofcontents

\begin{document}
\maketitle

\section{Motivation---symmetries of gravity}

\noindent The main observation motivating this line of investigation is that gravity---and models containing gravity such as supergravity---in particular situations exhibit global symmetries reaching beyond those obtained as isometries. 
Important examples are maximal supergravities, which after dimensional reduction of some of the dimensions exhibit global symmetries which are the continuous versions of string theory dualities: T-duality and U-duality \cite{Hull:1994ys}. Even pure gravity, when dimensionally reduced to 3 or fewer dimensions, exhibit duality  symmetries, these are the famous Ehlers \cite{Ehlers:1957zz} and Geroch \cite{Geroch:1972yt} symmetries.

Table \ref{DualityExamplesTable} lists symmetry groups after dimensional reduction for gravity in $4$ and $D$ dimensions, and for $D=11$ supergravity. The entries on the last line are not  global symmetries of the reduced models, but instances of the conjectured Belinskii--Khalatnikov--Lifshitz (BKL) symmetry 
\cite{BKL,Damour:2001sa,Damour:2002mp} close to a space-like singularity.

If one believes that the appearance of these symmetries is not ``accidental'' (whatever that means), there is good reason to search for a formulation of the theory in question where the symmetry is kind of ``built in''. This does not mean that the model is equipped with a manifest global symmetry, but that there is a {\it local} symmetry in the formulation, which is large enough. It amounts to a reconsideration (locally, so far) of the principles of gravity, replacing its $GL(n)$ group by a suitable Kac--Moody group $G$. The local symmetry is a version of diffeomorphisms based on $G$ instead of $GL(n)$. This is extended geometry.

Extended geometry shows that the actual local symmetries of gravitational theories are much larger than the traditional diffeomorphisms (together with gauge symmetries of other fields). Section \ref{ExtGeomSection} gives a brief review of the basics. In Sections 
\ref{THASection} and 
\ref{ComplexSection}, I sketch the systematic way to formulate a first order Batalin--Vilkovisky (BV) dynamics for extended geometries.

\begin{table}
\begin{center}
\begin{tabular}{r|c|c|c|c}
&$D=4$ gravity&Gravity in $D$ dim's&$D=11$ supergravity&\\ \hline
&&$\vdots$&$\vdots$&\\
&&$A_{n-1}$&$E_n$\\
&&$\vdots$&$\vdots$&\\
$d=4$&&$A_{D-5}$&$E_7$&\\
$3$&$A_1$&$A_{D-3}$&$E_8$&Ehlers\\
$2$&$A_1^+$&$A_{D-3}^+$&$E_8^+\simeq E_9$&Geroch\\
$1$&$A_1^{++}$&$A_{D-3}^{++}$&$E_8^{++}\simeq E_{10}$&BKL\\
\end{tabular}
\caption{Symmetry groups after dimensional reduction from $D$ to $d=D-n$ dimensions.\label{DualityExamplesTable}}
\end{center}
\end{table}

\section{Brief introduction to extended geometry\label{ExtGeomSection}}

\noindent The idea of extended geometry
\cite{Cederwall:2017fjm,Cederwall:2018aab,Cederwall:2019qnw,Cederwall:2019bai,Cederwall:2021xqi,Cederwall:2023xbj,Bossard:2024gry}
 is to provide a geometrisation of the duality symmetries.
This is achieved by extending the ``internal'' coordinates (the ones discarded on dimensional reduction) to
fill a module of a Kac--Moody group $G$ (with Lie algebra $\fg$).
The dynamical fields belong to a coset $(G\times\RR^+)/K(G)$, where $K(G)$ is an involutory subgroup of $G$ (usually the maximal compact subgroup) transforming local frames.
There may also be other (non-gravitational) fields, tranforming in some $\fg$-modules.
The unifying concept of extended geometry originates in the 
corresponding constructions for T-duality and U-duality, double geometry \cite{Tseytlin:1990va,Siegel:1993xq,Siegel:1993bj,Hitchin:2010qz,Hull:2004in,Hull:2006va,Hull:2009mi,Hohm:2010jy,Hohm:2010pp,Jeon:2012hp,Park:2013mpa,Berman:2014jba,Cederwall:2014kxa,Cederwall:2014opa,Cederwall:2016ukd}
and exceptional geometry
\cite{Hull:2007zu,Pacheco:2008ps,Hillmann:2009pp,Berman:2010is,Berman:2011pe,Coimbra:2011ky,Coimbra:2012af,Berman:2012vc,Park:2013gaj,Cederwall:2013naa,Cederwall:2013oaa,Aldazabal:2013mya,Hohm:2013pua,Blair:2013gqa,Hohm:2013vpa,Hohm:2013uia,Hohm:2014fxa,Cederwall:2015ica,Butter:2018bkl,Bossard:2017aae,Bossard:2018utw,Bossard:2019ksx},
respectively.

\begin{figure}
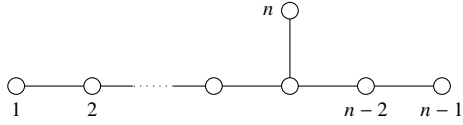

\begin{center}
\dynkin[backwards=true,make indefinite edge={5-6},labels={n-1,n,n-2,,,2,1}]E{ooooooo}
\caption{Dynkin diagram for $E_n$.\label{EnDynkin}}
\end{center}
\end{figure}

In our conventions, the coordinate module, or generalised vector module is a lowest weight $\fg$-module $R(-\lambda)$, where $\lambda$ is an integral dominant weight.
Examples are the U-duality groups $E_{n(n)}$, with $\lambda=\Lambda_1$ (Figure \ref{EnDynkin}),the Geroch group for $D=4$ gravity, which is the affine extension 
$A_1^+$ of $A_1\simeq SL(2)$, with $\lambda=\Lambda_1$ (Figure \ref{A1plusDynkin}), and the further over-extension $A_1^{++}$, the BKL group of $D=4$ gravity, with $\lambda=\Lambda_1$ (Figure\ref{A1plusplusDynkin}).

\begin{figure}
\begin{center}
\dynkin[arrows=false,labels={1,2}]B{oo}
\caption{Dynkin diagram for $A_1^+$.\label{A1plusDynkin}}
\end{center}
\end{figure}

\begin{figure}
\begin{center}
\dynkin[arrows=false,labels={1,2,3}]B{ooo}
\caption{Dynkin diagram for $A_1^{++}$.\label{A1plusplusDynkin}}
\end{center}
\end{figure}

We need to answer a couple of questions: What precisely are the local symmetries of the models? How is the original theory reproduced?

Denote a ``vector'', and element in $R(-\lambda)$ as $V^M$, so derivatives are $\*_M$, and let all fields and parameters depend on the extended coordinates $x^M$.
The local symmetries, the generalised diffeomorphsms, are generated by a generalised Lie derivative, which take the form
\begin{align}
\LL_\xi V^M=\xi^N\*_NV^M-\eta^{\alpha\beta}(t_\alpha\otimes t_\beta)_{NP}{}^{QM}\*_Q\xi^NV^P
+w(V)\*_N\xi^NV^M\;.\label{GenDiffEq}
\end{align}
Here, $\eta$ is the Cartan--Killing metric of $\fg$, and $t_\alpha$ are representation matrices of $\fg$ in $R(\pm\lambda)$.
$w(V)$ is the scaling weight of $V$, which for a (tensorial) vector turns out to be $(\lambda,\lambda)-1$.
Notice how the generalised Lie derivative mimics the ordinary one, with a transport term and a transformation term; the latter now implements a transformation in $\fg\oplus\RR$, replacing $\gl(n)$ as ``structure algebra''. By construction, the generalised Lie derivative obeys a Leibniz rule.
The action on (projections of) tensor products is defined by the Leibniz rule, so that the $t_\beta$ in eq. \eqref{GenDiffEq} is replaced by the appropriate representation matrix for the module in question. 
$\LL_\xi$ reproduces the ordinary Lie derivative when $\fg=\sl(n)$ and the Dorfmann bracket for $\fg=\so(n,n)$.

There are of course also ``ordinary'' diffeomorphisms (but dependent on the internal coordinates) for the $d$ external dimensions, as well
as other gauge symmetries.
They are straightforward to deal with, at least the procedures are known, so I focus on the ``internal'' dimensions, \ie, on the fields which in the dimensionally reduced theory are scalars.

Commuting two generalised diffeomorphisms returns a generalised diffeomorphism under some conditions/restrictions.
An important one is the {\it section constraint} on derivatives
\begin{align}
(-\eta^{\alpha\beta}t_\alpha\otimes t_\beta+(\lambda,\lambda)-1+\varsigma)\*\otimes\*=0.\label{SCEq}
\end{align}
The two derivatives can act on any pair of fields/parameters. $\varsigma$ is the permutation operator.
The so-called weak section constraint, which is eq. \eqref{SCEq} imposed on a single momentum, implies that momenta belong to the minimal $G$-orbit of $R(\lambda)$. The full constraint implies that momenta lie inside a linear subspace of this minimal orbit. 
There is a natural action of $GL(n)\subset G$ on such a subspace, a ``section''.
Solutions are found graphically in the Dynkin diagram by finding an $A_{n-1}$ diagram starting at the node dual to $\lambda$ (when it is a fundamental weight).
The r\^ole of the section constraint is thus to {\it locally} restrict the dependence on $x$ to the $n$ coordinates in the original (un-extended) theory. 
Locally, the picture of generalised geometry emerges, where space is not extended, but tangent space is.

The commutator of two generalised diffemorphisms then becomes
\begin{align}
[\LL_\xi,\LL_\eta]=\LL_{\lbr\xi,\eta\rbr}
\end{align}
in many relevant situations, including all physically relevant models when $d$ (the number of ``external'' dimension) is at least $4$ (for a full list
see ref. \cite{Cederwall:2017fjm}).
The ``Courant bracket'' $\lbr\xi,\eta\rbr$ is not a Lie bracket. In general, an $L_\infty$ algebra is obtained \cite{Cederwall:2018aab}.

In many interesting situations, in particular when $d\leq3$ and dual gravity is needed to fill the $\fg$-modules (Ehlers symmetry relates gravity and dual gravity), the situation is more complicated/interesting, and so called {\it ancillary} transformations appear. These are local $\fg$-rotations, with parameters that are constrained in a particular way related to the section constraint. 
I will not have time to discuss this in detail in this talk. Then, the commutator is
\begin{align}
[\LL_\xi,\LL_\eta]=\LL_{\lbr\xi,\eta\rbr}+\Sigma_{\xi,\eta}\;,
\end{align}
where the second term represents an ancillary transformation.

Actions, or rather pseudo-actions (given that the section constraint has to be imposed by hand) have been constructed in large classes of models, including exceptional geometry with structure group $E_{n(n)}$, $n\leq9$. 
Much of these constructions have been on a case by case basis. They have mostly employed a procedure where one starts from a generalised metric
parametrising the coset space $(G\times\RR)/K(G)$, but following a quite non-geometric path, where invariance under local $K(G)$ transformations is manifest, but invariance under generalised diffeomorphisms needs to be balanced between different terms in some Ansatz.

An alternative path, proposed in ref. \cite{Cederwall:2021xqi} and elaborated in ref. \cite{Cederwall:2023xbj}, leads to a formulation which is an improvement of the teleparallel formulation of gravity. The improvement actually implies a better control of the local $\so(n)$ transformations also in ordinary gravity. This formalism, which also has the advantage that we can predict the field content from an underlying superalgebra, will be described in the following Sections. 

\section{BV fields from tensor hierarchy algebras\label{THASection}}

\noindent Tensor hierarchy algebras \cite{Palmkvist:2013vya,Carbone:2018xqq,Cederwall:2019qnw,Cederwall:2021ymp,Cederwall:2023stz}
are a class of ``Cartan-type'' (roughly speaking containing vector fields) superalgebras,
the properties of which are precisely what is needed to provide an underlying structure encoding fields (including ghosts etc.) for extended geometry. 
Time does not allow for a detailed account, and I refer to \eg\ ref. \cite{Cederwall:2019qnw}.
The point is that a tensor hierarchy algebra $W(\fg)$ has a $\ZZ$-grading, where physical fields of extended geometry with structure algebra $\fg$ sit at degree 0, diffeomorphism ghosts at degree 1, higher ghosts at higher degree (in general, the algebra is infinite-dimensional, and there is no upper or lower limit to the degree). The elements at degree $-1$ can be identified with field strength (generalised torsion), at degree $-2$ with torsion Bianchi identities etc.
Letting all elements be functions of $x^M$ leads to a differential graded Lie algebra, \ie, there is a natural way of defining a nilpotent derivative carrying degree $-1$.
(Ancillary fields can be treated by an extension of this to a double grading of a tensor hierarchy algebra extension $S(\fg^+)$ of an extension $\fg^+$ of $\fg$, this is important but beyond the scope of this talk.) 
In addition, the generalised Lie derivative can be constructed as a derived bracket, using the bracket of the underlying tensor hierarchy algebra.

For the moment ignoring ancillary fields, this leads to a complex (the grading now goes to the left, degree equals ghost number) with the structure
\begin{equation}
    \begin{tikzcd}[row sep = 16 pt, column sep = 16 pt]
    \hbox{\hbox{\small gh\#=}}&\small 2&\small 1&\small 0&\small -1&\small -2&&\\
        \cdots\ar[r,"d"]&\hbox{\small $2^{\hbox{\tiny ary}}$ ghosts}\ar[r,"d"]&\hbox{\small diff. ghosts}\ar[r,"d"]&\hbox{\small vielbein}\ar[r,"d"]
        &\hbox{\small torsion}\ar[r,"d"]
        &\hbox{\small BI's}\ar[r,"d"]&\cdots \\
\end{tikzcd}
\label{THAComplex}
\end{equation}
However, extended geometry, like gravity, is a second order theory. We are not looking for torsion-free vielbeins.
How to achieve this from the present information is the subject of the next Section.

\section{Complexes and first order actions\label{ComplexSection}}

\noindent In a BV action, all fields have a conjugate antifield, with ghost numbers adding to $-1$. We need to use the complex \eqref{THAComplex}
together with its conjugate. In the conjugate complex, the derivative naturally acts in the opposite direction (think about exterior derivative and divergence). We still choose to call it $d$. The two lines thus obtained must be connected in some way by another part of the differential.

Lessons can be learned from (linearised, in this step) Yang--Mills theory in $d$ dimensions. 
Let us spend some time with this example, since it contains many relevant features of what we want to do later.
There, a complex $C_{\hbox{\tiny YM}}$:
\begin{equation}
\label{YMComplex}
    \begin{tikzcd}[row sep = 16 pt, column sep = 16 pt]
    \hbox{gh\#=}&1&0&-1&-2\\
       & \Omega^0\ar[r,"d"]&\Omega^1\ar[r,"d"]&\Omega^2 \\
        &&\Omega^{d-2}\ar[r,"d",swap]\ar[ur,"\sigma" near start]&\Omega^{d-1}\ar[r,"d",swap]
        &\Omega^d
\end{tikzcd}
\end{equation}
correctly reproduces the equations of motion.
The upper line contains the scalar ghost, the gauge connection and an antifield in the same module as the field strength.
By connecting it to the lower line of antifields (note that the one in $\Omega^{d-2}$ is a physical field) by an algebraic operator (\ie, one containing no derivatives), a first order formulation of Yang--Mills theory is obtained. 
Similar constructions have appeared in refs. \cite{Zeitlin:2008cc,Rocek:2017xsj,Reiterer:2019dys}.

The complex comes equipped with a natural BV pairing $\langle\cdot,\cdot\rangle$, carrying ghost number $1$, of fields with their antifields. 
The linearised BV action is 
\begin{align}
S=\fr2\langle\Psi,q\Psi\rangle\;,
\end{align}
where $q=d+\sigma$.

Let us check the action for the physical fields. Denote the physical 1-form in the upper line $A$ and the 2-form antifield $F$,
and their antifields $\bar A$ and $\bar F$. Note that $F$ is not the field strength of $A$, but an independent antifield, and that $\bar F$ is a physical field (ghost number 0).
We evaluate the relevant part of the action as
\begin{align}
S_0[A,\bar F]&=\fr2\left(
\langle A,d\bar F\rangle+\langle\bar F,dA+\sigma\bar F\rangle\right)\nn\\
&=\langle\bar F,F(A)\rangle+\fr2\langle\bar F,\sigma\bar F\rangle\;,
\end{align}
where $F(A)=dA$.
Solving the algebraic equations of motion obtained by varying $\bar F$ gives 
$\bar F=-\sigma^{-1} F(A)$, and reinserting in the action yields the standard two-derivative action
\begin{align}
S'_0[A]=-\fr2\langle F(A),\sigma^{-1}F(A)\rangle\;.
\end{align}

This elimination of $\bar F$ can equivalently be expressed as
homotopy transfer to the cohomology of $\sigma$.
This is done by first constructing a strong homotopy retract
to the cohomology of $\sigma$, which we call $C'_{\hbox{\tiny YM}}$, with zero differential,
\begin{equation}
    \begin{tikzcd}
(C'_{\hbox{\tiny YM}},0) \arrow[r, shift left=1ex, "{i}"] 
 &  \arrow[l, shift left=1ex, "{p}"]  (C_{\hbox{\tiny YM}},\sigma) \arrow[loop right, distance=3em, start anchor={[yshift=1ex]east}, end anchor={[yshift=-1ex]east}]{}{h}
\end{tikzcd}\;,
\end{equation}
where the inclusion $i$ and the projection $p$ are the na\"\i ve ones, identifying the spaces in $C'_{\hbox{\tiny YM}}$ with the ones in $C_{\hbox{\tiny YM}}$, and $h=\sigma^{-1}$.
Since $\sigma$ is invertible, the new complex $C'_{\hbox{\tiny YM}}$ consists of the vector spaces in $C_{\hbox{\tiny YM}}$, with $F\in\Omega^2$ and
$\bar F\in\Omega^{d-2}$ removed.
The horizontal differential $d$ is then seen as a perturbation of $\sigma$, and the homological perturbation lemma
\cite{Lapin} yields the quasi-isomorphism
\begin{equation}
    \begin{tikzcd}
(C'_{\hbox{\tiny YM}},q') \arrow[r, shift left=1ex, "{i'}"] 
 &  \arrow[l, shift left=1ex, "{p'}"]  (C_{\hbox{\tiny YM}},q=\sigma+d) \arrow[loop right, distance=3em, start anchor={[yshift=1ex]east}, end anchor={[yshift=-1ex]east}]{}{h'}
\end{tikzcd}\;,
\end{equation}
where the new differential $q'$ (and also $h'$, $i'$ and $p'$) is given as a perturbation series in $d$,
\begin{align}
q'=\sum_{n=0}^\infty p(dh)^ndi\;.
\end{align}
Here, only the terms with $n=0,1$ contribute, and dropping $p$ and $i$ (since they are given by the trivial identification of vector spaces), we obtain
\begin{align}
q'=d+d\sigma^{-1}d\;,
\end{align}
just like the result of algebraic elimination of fields above.
\begin{equation}
    \begin{tikzcd}[row sep = 16 pt, column sep = 16 pt]
        \Omega^0\ar[r,"d"]&\Omega^1\ar[dr,out=0,in=180,looseness=4,"d\sigma^{-1}d", near start]\\
        &&\bar\Omega^{d-1}\ar[r,"d",swap]
        &\bar\Omega^d
\end{tikzcd}
\end{equation}

The interacting YM theory is obtained by
``covariantisation'', which formally means Chern--Simons theory on the complex. The operator $\sigma$ of course remains undeformed.

Now, I would like to display the analogous construction for extended geometry (including ordinary gravity).
An important thing to notice is that there is a ghost for local rotations in $\fk$ (the Lie algebra of $K(G)$).
There must be an arrow without derivatives from this ghost to the vielbein, so the ghost is in the lower line.
This ghost must be paired with a ghost antifield (ghost number $-2$) in the upper line.
Fortunately, it turns out that the torsion Bianchi identity always contains the correct module.
(In ordinary gravity, the linearised Bianchi identity is
$(dT)_{mnp}{}^q=0$; contracting a lower and an upper index gives the index structure $B_{[mn]}$, which can be seen as an element of
$\so(n)$.)

We end up with the following complex $C$:
\begin{equation}\label{Cdiagram}
    \begin{tikzcd}[row sep = 16 pt, column sep = 16 pt]
    \hbox{gh\#=}&2&1&0&-1&-2&-3\\
        \cdots\ar[r,"d"]&V'\ar[r,"d"]&V\ar[r,"d"]&\hat\fg\ar[r,"d"]&\Theta\ar[r,"d"]&\bar\fk \\
        &&\fk\ar[r,"d",swap]\ar[ur,"\varrho" near start]&\bar\Theta\ar[r,"d",swap]\ar[ur,"\sigma" near start] 
        &\bar{\hat\fg}\ar[r,"d",swap]\ar[ur,"\varrho^\star" near start]&\bar V\ar[r,"d",swap]&\bar V'\ar[r,"d",swap]&\cdots
\end{tikzcd}
\end{equation}
Here, $\varrho$ is the standard embedding of $\fk$ in $\fg$ and $\varrho^\star$ the dual projection.
With the input from the tensor hierarchy algebra, the only unknown part of the differential is $\sigma$.
In order for $q^2=0$ where $q=d+\varrho+\sigma+\varrho^\star$, we need
\begin{align}
d\circ\varrho+\sigma\circ d=0\;.
\end{align}
This condition turns out to determine $\sigma$.

It is satisfying to verify that in ordinary gravity, the solution, acting on a tensor with all indices lowered, is
\begin{align}
(\sigma^{-1}X)_{m,np}=\fr2X_{m,np}-X_{[n,p]m}+2g_{m[n}g^{qr}X_{|q|,p]r}\;.
\end{align}
The three terms reproduce the standard three terms in the teleparallel formulation of gravity. 
I think the present formalism even sheds some light on this well known subject. In particular the r\^ole of the trace of the Bianchi identity as ghost antifield, and in that sense responsible for invariance under local Lorentz transformations is enlightening (checking this invariance of the standard teleparallel action produces a Bianchi identity; this is why).

I have so far considered linearised dynamics. The full non-linear dynamics is a matter of ``covariantisation''. One replaces the Lie algebra version of the vielbein by the full vielbein $E$, which is a group element. The arrow to torsion is deformed to taking the torsion of the Weitzenb\"ock connection of $E$ (a.k.a. the anholonomy coefficients). The algebraic operator $\sigma$ is undeformed. Some consistency check on the precise non-linear form of the Bianchi identities is needed, and turns out always to be fulfilled.

It should also be remembered that the above account was simplified in that it neglected ancillary ghosts and fields. 
They give rise to more algebraic arrows.

\section{Summary}

\noindent I have sketched how BV actions for extended geometry are constructed. In particular, given input from the relevant tensor hierarchy algebras, the main unknown is the map $\sigma$, whose inverse contracts two torsion tensors in the teleparallel action (in gravity, $\sigma^{-1}T$ is sometimes 
called the excitation form, see \eg\ ref. \cite{Krssak:2024vzo}). The problem is reduced to a purely algebraic one.
This is very useful in extended geometry, where one does not have the luxury of maintaining covariance with respect to both diffeomorphisms and local rotations in a Cartan formulation. This is due to the non-determinedness of the spin connection---unlike gravity, extended geometry does not have a matching of torsion and spin connection modules, and part of the spin connection remains undetermined by the vanishing of torsion.

This issue may actually point towards larger gauge algebras. One may introduce new ghosts to remove the undesired part of the spin connection, then consider the corresponding connection, which is in turn under-determined, and so on. This leads to an infinite-dimensional gauge algebra, which is known for double field theory
\cite{Butter:2021dtu}.
This whole procedure, as an extension downwards of the complex $C$, can in fact be viewed as unfolding of extended geometry (not on a solution of the section constraint, but on the extended space). This is work in progress with Falk Ha\ss ler. The complex $C$ can be obtained by homotopy transfer from the unfolding complex, which unfortunately is not  
equipped with a pairing. Hopefully, the construction of unfolding complexes can yield information that brings extended geometry closer to a true geometric  formulation. One example of a question that may be answered is: What is the proper generalisation of a form in extended geometry?

Of particular interest are situations where $\fg$ becomes infinite-dimensional. The present program has been implemented for
affine $\fg$ \cite{Bossard:2024gry}. The situation there is a bit atypical, probably due to the singular nature of affine Kac--Moody algebras. One needs an infinite tower of ancillary fields (labelled by increasing mode number) in order to write an action. I do not believe that this behaviour will persist for further extensions.
When $\fg$ is finite-dimensional, new modules start appearing in the tensor hierarchy algebras. For example, $\fg$ itself, at level 0, becomes extended by generators in some module of $\fg$. For affine $\fg$ it is a mode-shifted scalar, the Virasoro generator $L_1$. For over-extended $\fg$ it is an extension by generators in the fundamental module \cite{Cederwall:2021ymp}, and extra generators also appear at the level of vector fields etc. 
It was recently demonstrated \cite{Cederwall:2025muh} how this extension is related to  
the gradient structures appearing in the algebra \cite{Kleinschmidt:2005bq}, potentially opening for an algebraic emergence of space in the extended geometry framework.
In this context, it should be mentioned that an action based on $E_{11}$ has been proposed \cite{Bossard:2021ebg}. However, it does not use any of the ``extra'' fields described above, and should probably be considered a partial answer. It does provide a model locally equivalent to the bosonic sector of $D=11$ supergravity, and does in this sense provide a concrete realisation of the $E_{11}$ program (see \eg\ ref. \cite{West:2016xro}).

\bibliographystyle{utphysmod2}



\providecommand{\href}[2]{#2}\begingroup\raggedright\endgroup

\end{document}